\documentclass[twocolumn,american,aps]{revtex4}
\usepackage{lmodern}
\usepackage{lmodern}
\usepackage[T1]{fontenc}
\usepackage[latin9]{inputenc}
\usepackage[a4paper]{geometry}
\usepackage{color}
\usepackage{float}
\usepackage{amsmath}
\usepackage{amssymb}
\usepackage{graphicx}
\usepackage{soul}

\makeatletter

\providecommand{\tabularnewline}{\\}
\newcommand{\lyxdot}{.}

\@ifundefined{textcolor}{}
{%
 \definecolor{BLACK}{gray}{0}
 \definecolor{WHITE}{gray}{1}
 \definecolor{RED}{rgb}{1,0,0}
 \definecolor{GREEN}{rgb}{0,1,0}
 \definecolor{BLUE}{rgb}{0,0,1}
 \definecolor{CYAN}{cmyk}{1,0,0,0}
 \definecolor{MAGENTA}{cmyk}{0,1,0,0}
 \definecolor{YELLOW}{cmyk}{0,0,1,0}
}

\usepackage{babel}

\makeatother

\usepackage{babel}
\begin{document}

\title{Quantum Critical Exponents for a Disordered Three-Dimensional Weyl Node}
\begin{abstract}
Three-dimensional Dirac and Weyl semimetals exhibit a disorder-induced 
quantum phase transition between a semimetallic phase at weak disorder
and a diffusive-metallic phase at strong disorder. 
Despite considerable effort, both numerically and analytically, the 
critical exponents $\nu$ and $z$ of this phase transition are not
known precisely. Here we report a numerical calculation of the critical 
exponent $\nu=1.47\pm0.03$ using a minimal single-Weyl node model
and a finite-size scaling analysis of conductance. Our high-precision
numerical value for $\nu$ is incompatible with previous numerical 
studies on tight-binding models and with one- and two-loop calculations in an $\epsilon$-expansion
scheme. We further obtain $z=1.49\pm0.02$
from the scaling of the conductivity with chemical potential. 
\end{abstract}

\author{Bj\"orn Sbierski, Emil J. Bergholtz, and Piet W. Brouwer}

\affiliation{Dahlem Center for Complex Quantum Systems and Institut f\"ur Theoretische
Physik, Freie Universit\"at Berlin, D-14195, Berlin, Germany}

\date{September 9th, 2015}

\maketitle

\paragraph{Introduction.}
Materials with an electronic bandstructure dispersing linearly from
a Fermi point are among the driving themes in contemporary condensed
matter physics \cite{Burkov2015,Cayssol2013,Wehling2014}. 
After the experimental verification of such a 
Dirac-type bandstructure in single-layer graphene \cite{Novoselov2005},
the focus has now turned to three-dimensional materials. The compounds
Na$_{3}$Bi and Cd$_{3}$As$_{2}$ have been confirmed as ``Dirac semimetals''
\cite{Liu2014,Neupane2014,Borisenko2014,Jeon2014a,He2014}. 
In materials that break either time- or space inversion symmetry,
the twofold band degeneracy of Dirac semimetals is lifted and the
resulting phase is termed Weyl semimetal. The non-centrosymmetric
compounds TaAs and NbAs have recently proven experimentally to 
harbor such Weyl nodes in their bandstructures \cite{Lv2015,Xu2015,Xu2015b}. Similar bandstructures have been achieved in a photonic crystal realization in Ref. \cite{Lu2015}.

Theoretical work accompanied and, in part, preceded the recent
experiments. Beyond the single particle picture, Coulomb interactions 
were argued to be marginally irrelevant in the renormalization group 
(RG) sense due to the vanishing density of states at the Fermi point
\cite{Goswami2011,Hosur2012}.
Quenched disorder, however, inevitably present in realistic 
materials, is a much more subtle issue. 
Dating back to work from the 1980s \cite{Fradkin1986,Fradkin1986a},
the presence of a disorder-induced quantum phase transition is by
now firmly established 
analytically \cite{Goswami2011,Ominato2014,Syzranov2014,Syzranov2014a}
and numerically \cite{Kobayashi2014,Pixley2015,Pixley2015a, Bera2015, Liu2015,Sbierski2014a}. In
the weak-disorder phase, the random potential is irrelevant in
an RG sense. Thus, for large system sizes and low temperatures, a
weakly disordered system qualitatively behaves as a clean system with
renormalized Fermi velocity. This leads to a number
of experimentally important predictions for weak disorder, such as
quadratically vanishing density of states 
\cite{Kobayashi2014,Syzranov2014a} or pseudoballistic charge transport
\cite{Sbierski2014a} at the nodal point. In contrast, for strong disorder
one finds a metallic phase with finite
density of states at the Fermi energy and diffusive transport characteristics
\cite{Sbierski2014a,Altland2015}.

Signatures of the disorder-induced quantum criticality are expected
in almost any experimentally relevant observable, from heat capacity
to transport properties. Standard scaling theory \cite{CardyBook}
predicts power-law dependences on disorder, chemical potential, or 
temperature in the vicinity of the critical
point \cite{Kobayashi2014,Syzranov2014}. The only input to this variety
of predicted power laws is a pair of critical exponents characteristic
of the universality class. Denoting the dimensionless disorder strength and
chemical potential by $K$ and $\mu$, respectively, close to the
critical point $K=K_{\rm c}$, $\mu=0$ the \emph{correlation length exponent}
$\nu$ and the \emph{dynamical critical exponent} $z$ govern the
relation between reduced disorder strength $k=|K-K_{\rm c}|/K_{\rm c}$ and
the emerging correlation length $\zeta$ as $\zeta\propto k^{-\nu}$,
and the relation between emergent energy- and length scales as 
$\varepsilon\propto\zeta^{-z}$.
Although the critical point is located at zero chemical potential,
predictions of scaling theory persist for small finite doping.

To date, the best analytical estimates for $\nu$ and $z$ for a single Weyl- or Dirac node follow
from a Wilsonian momentum shell RG calculation in an $\epsilon$-expansion
scheme around critical dimension two. The results of the one-loop calculation
by Goswami and Chakravarty are $\nu=1$ and $z=1.5$ \cite{Goswami2011}.
The accuracy of the one-loop exponents was challenged by a calculation
of two-loop diagrams by Roy and Das Sarma \cite{Roy2014}, who found
$\nu=1.14$ and $z=1.31$. On the other hand, there are instances
where the $\epsilon$ expansion strategy is known to fail completely,
the Anderson metal-insulator transition in three dimensions being a well 
known example \cite{Kramer1993} --- although the present transition is of a 
different type as it connects two non-insulating phases 
\cite{Pixley2015}. Numerical results for the critical exponents obtained from tight-binding models harboring multiple Weyl- or Dirac nodes \cite{Kobayashi2014,Pixley2015,Pixley2015a, Bera2015, Liu2015} are in reasonable agreement with the one-loop results above, albeit with large uncertainties in $\nu$, $z$ and $K_\mathrm{c}$. 

Motivated by the lack of a firm theoretical prediction and in view of potential 
experiments, we performed a numerical calculation of the critical
exponents in a single Weyl node using state of the art finite-size scaling for quantum transport 
properties. Our results, which we report in detail below, have significantly 
reduced uncertainties in comparison to 
the previously known estimates. Whereas our result for the dynamical
critical exponent, $z =1.49\pm0.02$, is 
consistent with the previous numerical calculations and with the one-loop 
calculation (but not with the two-loop calculation!), our value for the
correlation length exponent, $\nu=1.47\pm0.03$, deviates rather
significantly.

\paragraph{Minimal model and numerical method.}

The minimal model for the disorder induced quantum criticality is
a \emph{single} Weyl node with potential disorder, 
\begin{equation}
H=\hbar v\boldsymbol{\sigma}\cdot\mathbf{k}+\mu+U(\mathbf{r}),\label{eq:H}
\end{equation}
where $v$ is the Fermi velocity, $\boldsymbol{\sigma}$ denotes the
vector of Pauli matrices, and $\mathbf{k}$ measures the Bloch wavevector
relative to the nodal point. We connect the Weyl semimetal to two 
ideal leads, both modeled as Weyl nodes with
$\mu$ taken 
to infinity and without the random potential $U$.
The Weyl semimetal has dimension $0< x < L$ and $0 < y,z < W$ in 
transport and transverse directions, respectively. 
To quantize transverse momenta $k_{y,z}$ we apply periodic or antiperiodic
boundary conditions (PBC/APBC). 
An ultraviolet cutoff $\Lambda$ restricts the
magnitude of transverse wavevector $|k_{y,z}| \leq
\Lambda$ and $1/\Lambda$ sets the microscopic length scale. The 
random potential $U\left(\mathbf{r}\right)$
is assumed to have zero mean and Gaussian white noise fluctuations
\begin{equation}
\left\langle U(\mathbf{r})U(\mathbf{r}^{\prime})\right\rangle=\frac{K}{\Lambda}(\hbar v)^{2}\delta(\mathbf{r}-\mathbf{r}^{\prime}),\label{eq:disCorr}
\end{equation}
with $K$ the dimensionless disorder strength. The chemical potential
$\mu$ has to vanish to reach the critical point, however, we will
work with finite $\mu$ below to assess the dynamical critical exponent
$z$.

We study the signatures of disorder-induced quantum criticality of
Eq. (\ref{eq:H}) in a quantum transport framework at zero temperature,
employing the numerical scattering matrix method of Ref.\ 
\cite{Sbierski2014a},
which is based on related studies of disordered Dirac fermions in
two dimensions \cite{Bardarson2007,Adam2009}. 
The conductance can be computed
from the scattering matrix's transmission block $t$ using the Landauer
formula $G=\mathrm{tr}\, tt^{\dagger}$
and is measured in units of $e^{2}/h$ throughout.

\paragraph{Correlation length exponent $\nu$: Finite-size scaling of the conductance for $\mu=0$.}

The standard method to assess the
correlation length exponent $\nu$ is finite-size scaling \cite{CardyBook}.
To perform such an analysis, one needs to identify a dimensionless
observable that assumes different values on the two sides of the (bulk)
phase transition. In Ref. \cite{Sbierski2014a} we showed numerically
that the conductance $G$ fulfills these requirements.
For large aspect ratio $r \equiv W/L \gg 1$ and in the thermodynamic
limit $L\rightarrow\infty$, the conductance 
takes the values $G_{K=0} = r^2 (\ln 2/2\pi)
\simeq 0.11 r^2$ in the pseudoballistic phase at disorder strength $K=0$ \cite{Baireuther2014}
and $G_{K \to \infty}=\sigma r^2 L\rightarrow\infty$ in the diffusive 
phase for $K > K_{\rm c}$ (with bulk conductivity $\sigma$)
\cite{Sbierski2014a}. 
In the vicinity of the critical point $K = K_{\rm c}$, when the system 
dimensions $L$, $W$ are larger than all internal length scales other than the
emerging correlation length $\zeta$, $G$ assumes a scaling form 
$G=G(L/\zeta,W/\zeta)$. Using $\zeta\propto k^{-\nu}$ and fixing
the aspect ratio $r = W/L$, we arrive at $G= G_{r}(L^{1/\nu}k)$,
with universal correlation length exponent $\nu$ and a
scaling function $G_{r}$ that depends on $r$ and the boundary 
conditions.

Numerically, the conductance $G$ is found to vary considerably
for different disorder realizations, however with the restriction
$G>G_{K=0}$ for every disorder realization. The histogram of $\delta G \equiv
G-G_{K=0}$
is shown in Fig. \ref{fig:PDF} for the specific choice $L=2\pi/(\Lambda r)\times11$,
PBC. For weak disorder, $K\lesssim5$, we find that the distribution
of $\delta G$ can be well fitted by a log-normal distribution 
$p_{\rm LN}(\delta G)=e^{-(\ln \delta G -\mu_{\rm LN})^{2}/2\sigma_{\rm LN}^{2}}/\sigma_{LN}\delta G\sqrt{2\pi}$,
with parameters $\mu_{\rm LN}$ and $\sigma_{\rm LN}$, see 
Fig.\ \ref{fig:PDF}.
A feature not captured by this fit is the tail of large but rare 
conductances, which are possibly related to rare region effects 
\cite{Nandkishore2014}. As $G$ is not self-averaging (analogous to 
Anderson localization), we choose the median $m$ of the distribution 
as a scaling quantity, {\em i.e.}, we search a scaling function
$m=m_{r}(L^{1/\nu}k)$. The standard error
of the median is calculated using the asymptotic variance formula
$\sigma_{m}^{2}=1/4p(m)^{2}N$, where $N$
is the total number of disorder realizations and the unknown exact
probability density $p$ is approximated by a smooth interpolation
of the measured histogram. For further research, it would be desirable
to understand the occurrence of the empirical log-normal conductance
distribution.
\noindent \begin{center}
\begin{figure}[H]
\begin{centering}
\includegraphics[scale=0.93]{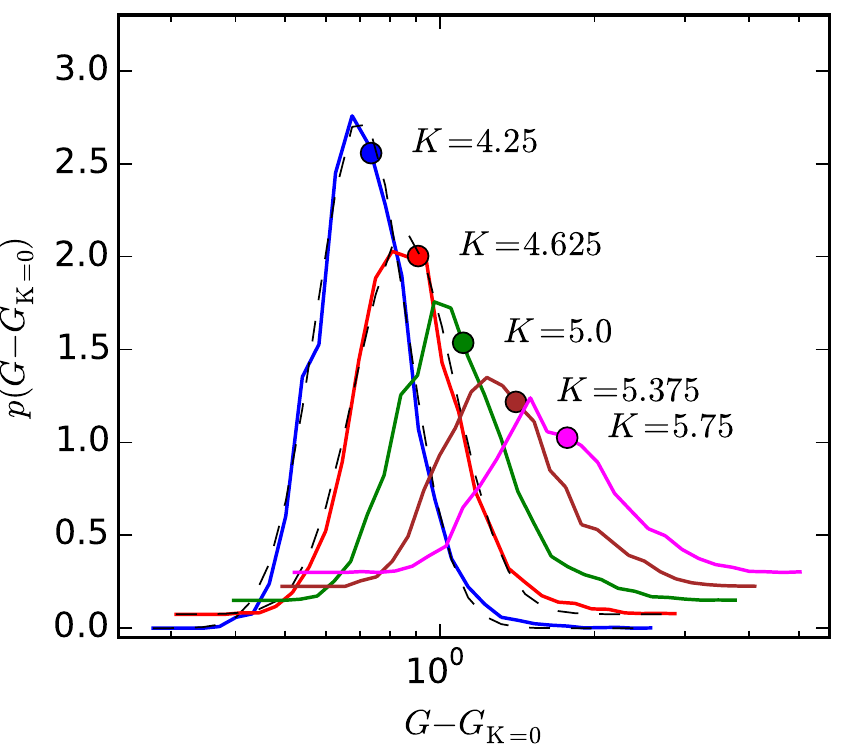} 
\par\end{centering}
\protect\caption{{\small{}\label{fig:PDF}(color online) 
Probability distribution of the difference $\delta G = G - G_{K=0}$ of 
the conductance with and without disorder potential. The aspect ratio
$r = W/L=5$, periodic boundary conditions were applied in the transverse
direction, and the sample length $L=(2\pi/\Lambda)\times 11/r$. 
The dimensionless disorder strength is indicated in the figure.
The number of disorder realizations is 4000. Dots indicate the value of 
the median $m$, which is used as the scaling variable. Dashed lines show 
fits to a log-normal probability distribution. The data for $K>4.25$ is 
offset in vertical direction for clarity.}}
\end{figure}
\par\end{center}

We compute $m$ for a range of disorder strengths $K$ and lengths
$L$, for aspect ratios $r=5$ and $7$, and we also varied the boundary 
condition between periodic and anti-periodic. The data for $r=5$,
PBC, is shown in Fig.\ \ref{fig:scaling}, the other data sets
can be found in the supplemental information, Ref. \cite{SI}. At criticality, where $\zeta$
diverges, $m_{r}$ is independent of $L$ and the data traces in Fig.
\ref{fig:scaling} cross in one point. In the supplemental information \cite{SI}, we show the
details of a least squares fit for $m_{r}\left(L^{1/\nu}k\right)$
for small $k$ to a polynomial of fourth order in $L^{1/\nu}k$ (solid
lines). An excellent and stable fit was achieved even without including
any irrelevant scaling variable that we took in leading order as $L^{y}$ 
with $y<0$. Taking into account the fitting results of all other
parameter sets in a standard procedure (see \cite{SI}) we find
$\nu=1.47\pm0.03$.
The conductance data for smaller aspect ratios $r \lesssim 3$
(data not shown) reveals a large irrelevant contribution to the 
scaling function that hindered
a successful fit in terms of a simple low order polynomial.

In Ref. \cite{Sbierski2014a} it was argued that the Fano factor $F$ (the ratio of shot-noise power and conductance) is an alternative quantity to distinguish the pseudoballistic from the diffusive phase. In the pseudoballistic phase one has $F(K<K_\mathrm{c})\simeq0.574$ while in the diffusive phase $F(K>K_\mathrm{c})=1/3$. In the Supplemental Material we show that our result for $\nu$ is consistent with the value obtained from a finite-size scaling analysis using the Fano factor ($\nu=1.40\pm0.05$). This method however suffers from an inferior quality of the data set, both in terms of error bars of the individual data points as well as in the range of system sizes available.
\noindent \begin{center}
\begin{figure}
\begin{centering}
\includegraphics{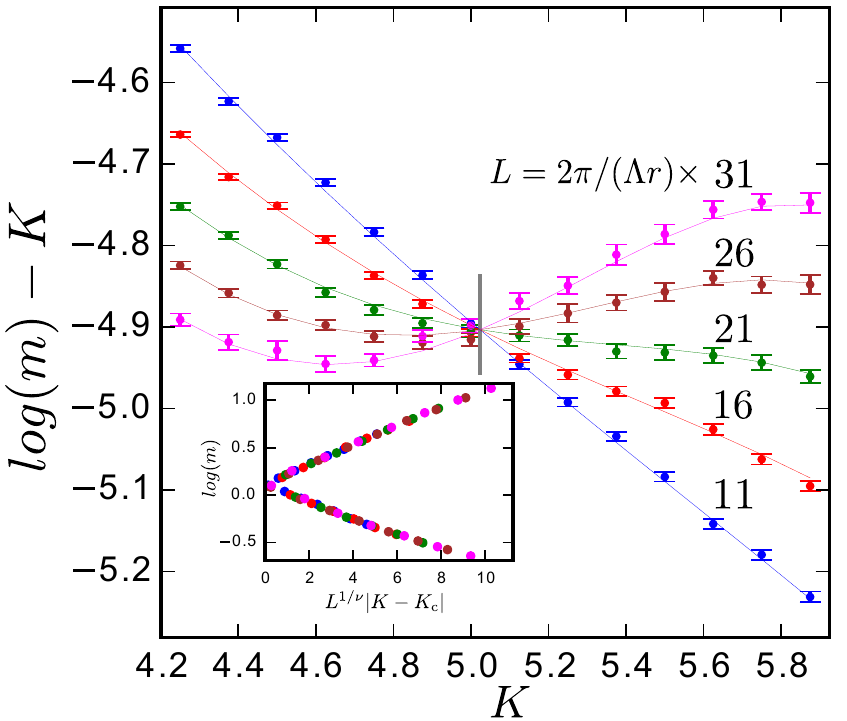} 
\par\end{centering}
\protect\caption{{\small{}\label{fig:scaling}(color online) 
Scaling plot for the logarithm
of the median $m$ of the distribution of $\delta G = G - G_{K=0}$
for a disordered Weyl node with $W/L=5$ and periodic transverse
boundary conditions. For better visibility we plot $log(m)-K$ vs. $K$
on the vertical axis. The solid curves show the results of a 
least-squares fit to a scaling form $m=m_{r}(L^{1/\nu}k)$, 
expanded in a fourth order polynomial (for details, see \cite{SI}).
The estimates for the most important fit parameters and their standard
error are $K_{\rm c} = 5.024 \pm 0.004$, $\log(m_{\rm c}) = 0.12 \pm 0.004$,
$\nu = 1.47 \pm 0.02$. The quality of the fit is $\chi^2/N=0.93$.
The sample lengths are indicated in the figure.
The gray vertical line indicates the position of the estimated
critical disorder strength. The inset shows a scaling collapse of the data in the main panel using the estimated values of $K_{\rm c}$ and $\nu$. }}
\end{figure}
\par\end{center}

\paragraph{Dynamical critical exponent: Scaling of critical bulk conductivity.}
We now turn to the dynamical critical exponent $z$ that connects
the emergent length scale $\zeta$ and the corresponding energy scale
$\varepsilon$ in the vicinity of the fixed point. 
In our transport geometry, a natural choice of a quantity that
has a scaling involving the dynamical exponent $z$ is the 
bulk conductivity $\sigma$,
which is also of immediate experimental relevance.

To connect the dynamical critical exponent to the conductivity, we
again start with a scaling form around criticality \cite{Syzranov2014}.
Since the unit of $\sigma$ in three dimensions is inverse length, we find $\sigma\left(k,\mu\right)=\zeta^{-1}f(\mu/\varepsilon)$
  with an unknown dimensionless scaling function $f$. We define a new scaling function $f\left(x\right)=x^{1/z}\tilde{f}\left(x^{-1}\right)$
  in terms of which $\sigma\left(k,\mu\right)=\mu^{1/z}\tilde{f}\left(k^{z\nu}/\mu\right)$
 . At $K=K_{\mathrm{c}}$, the `critical' conductivity $\sigma_{\mathrm{c}}$ thus scales as
\begin{equation} 
\sigma_{\mathrm{c}}\left(\mu\right)\propto\mu^{1/z}.
\label{eq:critical_conductivity_scaling}
\end{equation}
 The scaling form (\ref{eq:critical_conductivity_scaling}) is valid with small corrections within an extended quantum critical region \cite{Syzranov2014a} for finite $k$ when the argument of $\tilde{f}$
  is sufficiently small, \emph{i.e.} $k\ll k^{*}\left(\mu\right)\propto\mu^{1/(\nu z)}$.
This allows us to numerically compute an estimate of $z$ in spite of
the fact that the value of $K_{\rm c}$ is known only within error bars.

We compute $G(L)$ for fixed large $W$, PBC, a range of $\mu$ and
for $K=5.0$, which is within the $K_{\rm c}$ confidence interval \cite{SI}. We perform a
disorder average over at least ten disorder realizations. Since 
transport in a Weyl node at finite $\mu$ is diffusive,
we expect $G = \sigma W^2/L$, which is confirmed in the simulation.
Finite size effects are irrelevant once $W,L$ are larger than the 
characteristic $\mu$-induced length scale $\propto\mu^{-1/z}$. 
We show $\sigma_{c}$ vs.\ $\mu$ for $K = K_{\rm c}$ in Fig.\ 
\ref{eq:critical_conductivity_scaling} (dots) and indeed
observe a power law (solid line) for $\mu\lesssim\hbar v\Lambda$
with inverse exponent $z=1.49\pm0.02$. For larger chemical
potentials, $\mu$ comparable to the band edge $\hbar v\Lambda$, the scaling
breaks down and from Drude transport theory we expect a crossover
to $\sigma \propto \mu^2$, proportional to the density of states (dashed 
line). 
\noindent \begin{center}
\begin{figure}
\noindent \begin{centering}
\includegraphics[scale=0.81]{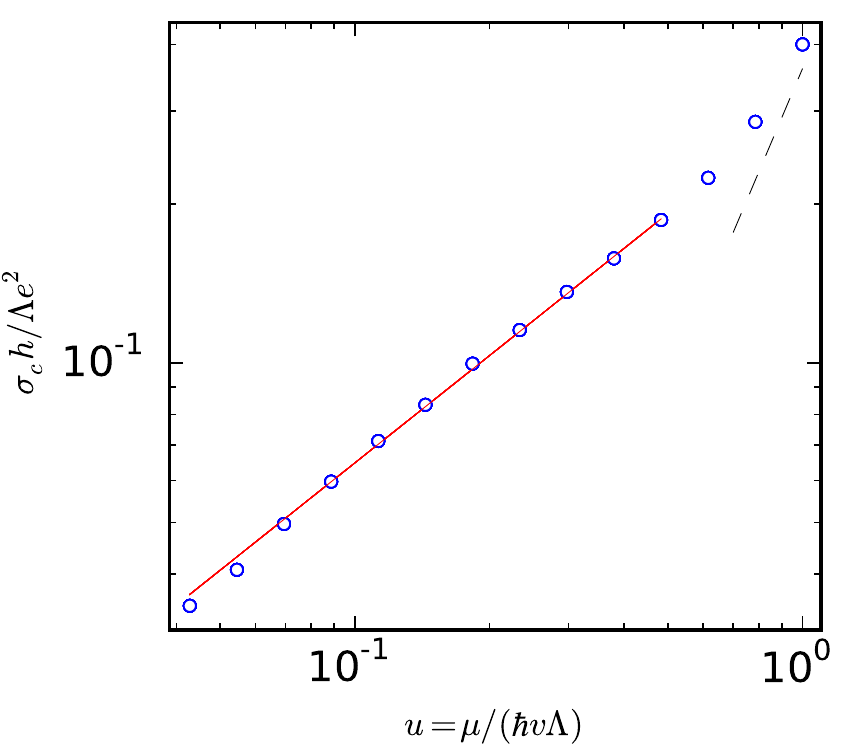} 
\par\end{centering}
\protect\caption{{\small{}\label{fig:sigma-u-scaling}
Scaling of the conductivity $\sigma\left(K,\mu\right)$ for (near) 
critical disorder strengths
$K=5.0$ (dots) with chemical potential $\mu$. The solid line 
is a power-law fit $\sigma \propto \mu^{1/z}$ to the data points, with
$z = 1.49 \pm 0.02$. The dashed line indicates a $\sigma\propto\mu^2$ power law as expected from Drude transport theory when scaling breaks down.
The sample width $W = (2 \pi/\Lambda) \times 29$ and periodic 
boundary conditions were applied.
For chemical potentials below the values 
shown in the figure a bulk conductivity could not be reliably obtained
from the calculated conductance data.}}
\end{figure}
\par\end{center}
\paragraph{Discussion.}
We numerically studied the disorder-induced 
quantum phase transition in three-dimensional Dirac materials in terms of a minimal model, a single Weyl node with potential disorder. In contrast to the well-known Anderson metal-insulator
transition, this disorder-induced phase transition for a single Weyl node is between two
non-insulating phases. In addition to the correlation-length 
exponent $\nu$, it features a non-trivial 
dynamical critical exponent $z$, which has no counterpart in the 
standard metal-insulator transition. 

Our high-precision results for the exponents $\nu$ and $z$ not only 
allow for a variety of quantitative
predictions of experimentally observable power laws around criticality
--- such as the density-of-states exponent $\beta = \nu(3-z)$ for
$K > K_{\rm c}$ \cite{Kobayashi2014} ---
but also improve on previously reported theoretical predictions. Our
result $\nu=1.47\pm0.03$ differs significantly 
from analytical results 
obtained from a one- or two-loop $\varepsilon$-expansion RG calculation 
($\nu=1$, $1.14$, respectively \cite{Goswami2011,Roy2014}). 
The failure of the $\epsilon$-expansion calculation
is reminiscent of the situation for the Anderson localization in three
dimensions \cite{Kramer1993}, where $\nu=1.375$ in the symplectic class \cite{AsadaY2005}.
Our estimate for the dynamical critical exponent is $z=1.49\pm0.02$,
in agreement with the one-loop RG calculation ($z=1.5$), but not with the 
two-loop prediction ($z=1.31$). 

In principle, the model of Eq. (\ref{eq:H}), which has white-noise 
disorder and a sharp momentum cutoff, could be modified to include a
more faithful representation of the microscopic disorder, albeit at
an increased numerical cost. For example, Ref.\ \cite{Sbierski2014a}
employed a finite disorder correlation length that sets
the microscopic length scale; the mode cutoff then can safely be taken
to infinity. However, the difference between these two models is irrelevant
in the RG sense and thus both models are in the same universality class. 
To see this, recall that a finite disorder correlation length
is equivalent to a higher-order momentum dependence of the 
disorder-induced interaction vertex in the replicated disorder averaged 
action and is thus irrelevant \cite{Shankar1994}. On the other hand, 
the numerical
value of $K_{\rm c}$ is non-universal and, thus, sensitive to the disorder
model. In this context we note that a model with sharp
momentum cut-off has also been used in Ref.\ \cite{Bardarson2007}, 
where it was found to give the same results as a model with finite
disorder correlation length.
Moreover, in a realistic band structure the linear form
of Eq. (\ref{eq:H}) is only an approximation. Quadratic corrections,
however, are RG irrelevant and thus will not change the critical exponents
\cite{Goswami2011}. 
\paragraph{Realistic Weyl and Dirac semimetals.}
Realistic Weyl or Dirac semimetals have multiple Weyl nodes \cite{Nielsen1981}, either separated in momentum space or distinguished by their transformation properties under point group symmetries. The same applies to numerical studies based on tight-binding models \cite{Kobayashi2014, Pixley2015, Pixley2015a, Bera2015, Liu2015}, which confirmed the presence of a disorder-induced phase transition on the basis of density-of-states calculations. With multiple Weyl nodes, disorder might not only cause intra- but also inter-node scattering of quasiparticles. The latter process is not captured by our minimal model. Symmetries in more realistic models with multiple Weyl nodes may also
be different from the minimal model: While our minimal model has an
effective time-reversal symmetry mapping the single Weyl node onto
itself, in realistic models, time-reversal or inversion symmetries can
relate different nodes or be absent.
Although the precise nature of disorder potentials in realistic three-dimensional Dirac materials is yet to be determined, there are plausible scenarios in which intra-node scattering dominates over inter-node scattering, a priori justifying the use of our minimal model. For example, in Weyl semimetals the ratio of scattering rates is controlled by the smoothness of the disorder potential and the separation of Weyl nodes in momentum space \cite{footnote_Dirac}.

The applicability of the minimal model in the presence of sizeable inter-node scattering, {\em i.e.}, the question whether or not the presence of some amount of inter-node scattering changes the universality class of the disorder-induced semimetal-metal transition, is an issue that has not been conclusively settled \cite{footnote_AL}. Inter-node scattering is omitted in field-theoretical approaches \cite{Goswami2011,Syzranov2014}. Empirical evidence that inter-node scattering does not affect the universality of the transition comes from Ref.\ \cite{Pixley2015a}, which found a remarkable universality for three different disorder types in a tight-binding Dirac semimetal model, albeit with large error bars on the critical exponents.


If the assertion of a single universality class insensitive to inter- or intra-cone scattering (and the related symmetry differences) is correct, the observed critical exponents should match with those obtained in tight-binding models. However, such studies \cite{Kobayashi2014, Pixley2015, Pixley2015a, Bera2015, Liu2015} yield a value for $\nu$ inconsistent with our result, for a typical example see Ref. \cite{Kobayashi2014} which finds $\nu=0.92\pm0.13$ at $K > K_{\rm c}$ and $\nu=0.81\pm0.21$ at $K < K_{\rm c}$. The value of the dynamical critical exponent $z=1.5\pm0.1$ from Ref. \cite{Kobayashi2014} is in agreement with our result.
Assuming that the type of scattering is indeed immaterial for critical exponents, we attribute the large difference with the tight-binding model exponent $\nu$ to difficulties in accurately
estimating the critical disorder strength from density-of-states data. The uncertainty of $K_{\rm c}$ translates to a 
large uncertainty in the critical exponent $\nu$. In contrast,
our very precise estimate of $K_{\rm c}$ was possible using the 
finite-size scaling method where $K_{\rm c}$ can be obtained from the unique crossing of the data in Fig. \ref{fig:scaling}. 

In the supplemental information \cite{SI}, we exemplify this interpretation by revisiting the density of states data from Ref. \cite{Kobayashi2014} (cf. Fig. 3a) obtained at zero energy for a range of disorder values around the critical disorder strength. Using the critical exponent $\beta=2.22$ calculated with our estimates for $\nu$ and $z$ we are able to produce an excellent fit for the data points in the vicinity of the critical disorder strength, though we find a much smaller critical disorder strength than asserted in Ref. \cite{Kobayashi2014}. Since the microscopic model in Ref. \cite{Kobayashi2014} and in this work are different, the values of the non-universal critical disorder strengths cannot be compared. 
Uncertainty in $K_{\rm c}$ does not cause comparable problems when
determining the critical exponent $z$, because the large size of
the critical region in the chemical potential--disorder parameter 
plane renders the extraction of $z$ much less sensitive to the 
uncertainty in $K_{\rm c}$.
This is consistent with the mutual agreement between our estimate
for $z$ and the value in Ref.\ \cite{Kobayashi2014}.

\paragraph{Acknowledgments.}

It is a pleasure to thank Johannes Reuther, Maximilian Trescher and Tomi Ohtsuki for helpful discussions
and J\"org Behrmann and Jens Dreger for support on the computations
done on the HPC cluster of Fachbereich Physik at FU Berlin. We further acknowledge discussion with Sankar Das Sarma and correspondence with Pallab Goswami and Jed Pixley. Financial
support was granted by the Helmholtz Virtual Institute ``New states
of matter and their excitations'', by the Alexander von Humboldt
Foundation in the framework of the Alexander von Humboldt Professorship,
endowed by the Federal Ministry of Education and Research, and by the
DFG's Emmy Noether program (BE 5233/1-1).


\begin{references}
\bibitem{footnote_Dirac} For 
disordered Dirac semimetals where Weyl nodes coincide, spatial symmetries
of the specific model can suppress inter-node scattering.
\end{references}

\clearpage{}

\onecolumngrid

\section*{Supplemental Material}
\paragraph{Details of the finite-size scaling analysis.}

We here provide details of the finite-size
scaling procedure, following Refs.\ \cite{Slevin1999,Obuse2012,Obuse2014,Obuse2013a}.
In addition to the data set presented in the main text --- aspect ratio
$r=5$ and periodic boundary conditions (PBC) ---, we have obtained
conductance distributions for antiperiodic boundary conditions (APBC)
with $r=5$ and for aspect ratio $r=7$, PBC.

The sample width $W=L\cdot r$ is set to be $W = (2 \pi/\Lambda) (M-1/2)$ for APBC
and $W = (2 \pi/\Lambda) M$ for PBC, with $M$ a positive integer. The 
transverse wavenumbers are $k_{y,z} = (2 \pi/W) n_{y,z}$, with $n_{y,z}
= -M, -M+1,\ldots,M$ for PBC and $n_{y,z} = -M+1/2,-M+3/2,\ldots,M-1/2$
for APBC. A summary of all data
sets used in this work is given in Table \ref{tab:Details-of-the-FSS}.

\begin{figure}[H]
\begin{centering}
\includegraphics{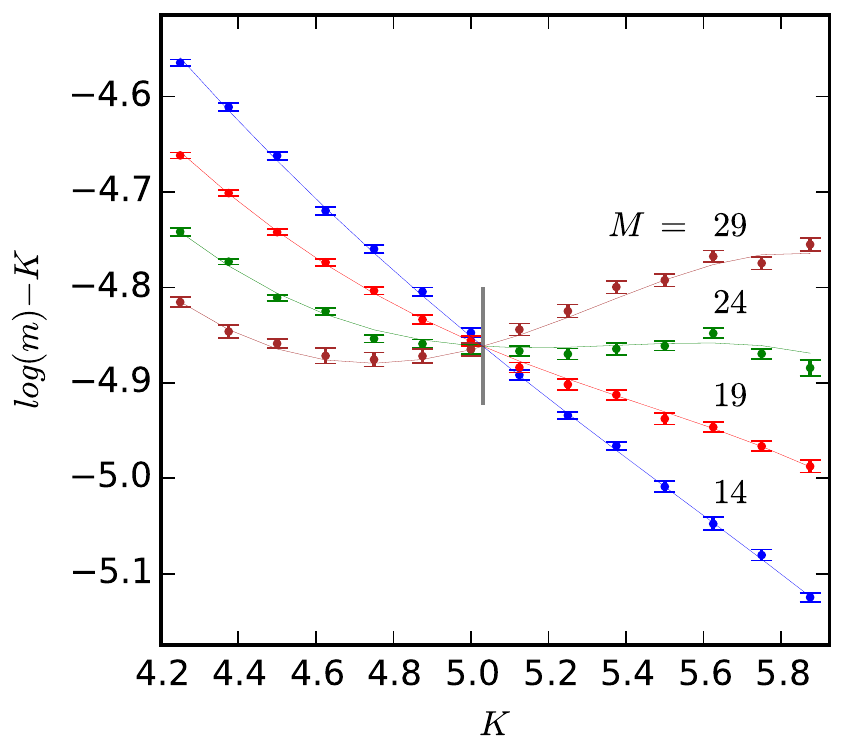}\includegraphics{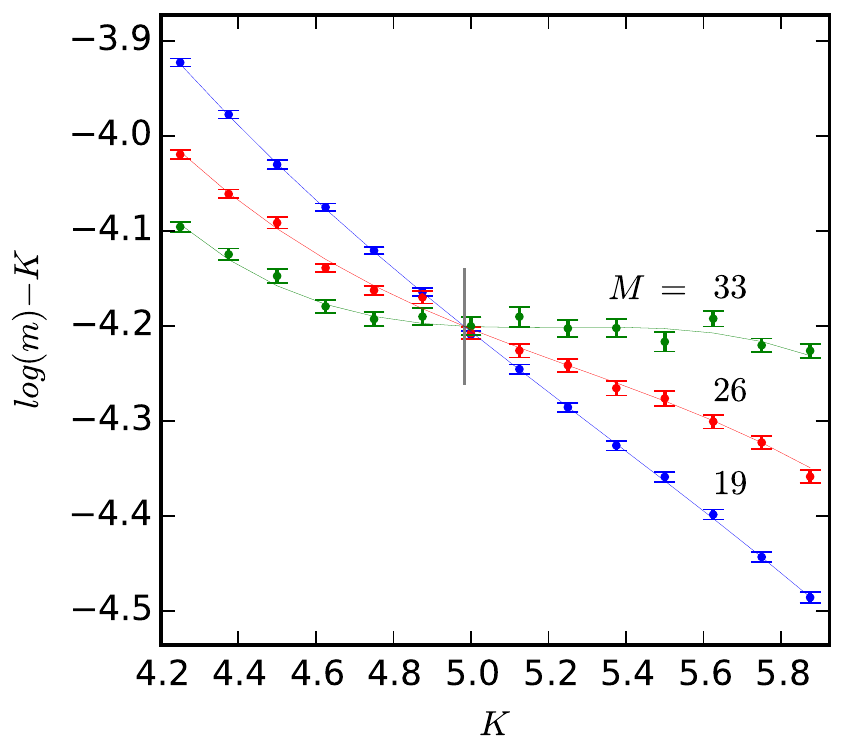}
\par\end{centering}

\protect\caption{{\small{}\label{fig:additional scaling plot}(color online) 
Scaling
plots for the logarithm of the median $m$ of the distribution of the
difference $\delta G = G - G_{K=0}$ of the two-terminal conductance
with and without disorder potential, for a Weyl node with aspect ratio
$r=W/L=5$ and antiperiodic boundarity conditions (left) and $r=7$ and periodic boundary conditions (right). For clarity, we plot $log(m)-K$ vs.\ $K$. The
solid curves show the results of a least-squares fit to
a scaling form $m=m_{r}\left(L^{1/\nu}k\right)$ expanded in a fourth ($r=5$) and third ($r=7$) order polynomial, respectively. The
estimates for the most important fit parameters are given in 
Table \ref{tab:Details-of-the-FSS}. The gray vertical
line indicates the position of the estimated critical disorder strength.}}
\end{figure}

For each data set, the median $m(K,M)$ of the conductance distribution
is determined. We perform a least-squares fit to a polynomial of the form 
\begin{eqnarray}
  m(K,M) & = & a_{0}(1+b_{01}L^{y}+ \ldots + b_{0q_{0}}L^{yq_{0}})+a_{1}\cdot(L^{1/\nu}k)\cdot(1+b_{11}L^{y}+ \ldots + b_{1q_{1}}L^{yq_{1}})\label{eq:model}\\
 &  & \mbox{} + \ldots +a_{p}\cdot(L^{1/\nu}k)^{p}\cdot(1+b_{11}L^{y}+...b_{1q_{p}}L^{yq_{p}})\nonumber 
\end{eqnarray}
for medians obtained at the same value of the aspect ratio $r$ and the same boundary conditions. Data points ({\em i.e.}, medians of conductance distributions) and fits are shown in Fig.\ \ref{fig:scaling} of the main text for $r = 5$ and PBC, and in Fig. \ref{fig:additional scaling plot} for $r=5$, APBC, and and $r=7$, PBC. The following algorithm for the fitting procedure is used: The order of the polynomials in Eq. \eqref{eq:model}
is increased by adding a new fit parameter $a_{i}$ or $b_{ij}$ if
(i) the merit function $\chi^{2}/N\in[0,\infty]$ ($N$ is the number
of data points) for the resulting fit is lowered by more than 2\%
compared to the previous fit and (ii) the error of any fitting parameter
(as calculated from error propagation theory) does not exceed the
parameter's estimate in magnitude. Initial values for each fitting
procedure are chosen randomly and the parameter estimates for the
best fit out of a few hundred fitting trials is reported in Table
\ref{tab:Details-of-the-FSS} along with the error estimates and the
value of $\chi^{2}/N$. A fit is acceptable if $\chi^{2}/N<1$, another
measure is the `goodness of the fit' $\mathcal{G}\in[0,1]$ where
$\mathcal{G}=1$ indicates a perfect fit (for definitions of $\mathcal{G}$ and $\chi^{2}/N$ see, for example, Ref. \cite{Obuse2014}). 

Ideally, fitting parameters should not strongly depend on the number
of different values of $M$ within a data set. We successfully checked
the stability of the fitting results by repeating the fitting procedure
above for reduced data sets (deleting data points of the largest or
smallest $M$ in the data set $r=5$, PBC), as indicated in Table
\ref{tab:Details-of-the-FSS}. Finally, the estimate for $\nu$ is
calculated as an average of the best fit estimates for $\nu$ for
each data set whereas the total error bars are unions of error bars
from each single data set ('practical-error-bar procedure', see Ref.
\cite{Obuse2013a}).

\begin{center}
\begin{table}[H]
\begin{centering}
\begin{tabular}{|c|c|c|c||c|c|c|c|c|c|c|c|c|}
\hline 
$r=W/L$  & B.C.  & $M$  & $N$  & $\chi^{2}/N$  & $\mathcal{G}$  & $\nu$  & $K_{\rm c}$  & $log(m_\mathrm{c})=a_{0}$  & $a_{1} \cdot 10^1$  & $a_{2} \cdot 10^2$  & $-a_{3} \cdot 10^2$ & $-a_{4} \cdot 10^3$ \tabularnewline
\hline 
\hline 
5 & PBC   & $11,16,21,26,31$  & 70  & 0.93 &  0.4 & $1.47\pm0.02$ & $5.024\pm0.004$ & $0.12\pm0.004$ & $5.0\pm0.1$ & $3.0\pm0.3$ & $1.3\pm0.1$ & $1.8\pm0.9$    \tabularnewline
\hline 
5 & PBC   & $16,21,26,31$  	& 56  & 0.62 & 0.94  & $1.46\pm0.02$ & $5.007\pm0.007$ & $0.102\pm0.006$ & $4.9\pm0.2$ & $3.3\pm0.4$ & $1.2\pm0.1$ & $2.6\pm1.2$   \tabularnewline
\hline 
5 & PBC 	  & $11,16,21,26$  	& 56  & 0.84 &  0.55 & $1.48\pm0.02$ & $5.036\pm0.006$ & $0.128\pm0.004$ & $5.0\pm0.1$  & $3.2\pm0.4$ & $1.4\pm0.1$ & $2.3\pm1.3$   \tabularnewline
\hline 
5 & APBC & $14,19,24,29$  	& 56  & 0.97 &  0.29 & $1.47\pm0.02$ & $5.031\pm0.005$ & $0.169\pm0.004$ & $4.9\pm0.1$ & $2.8\pm0.3$ & $1.1\pm0.1$ & $1.7\pm0.8$    \tabularnewline
\hline 
7  & PBC  & $19,26,33$       	& 42  & 0.64 & 0.87 & $1.47\pm0.03$ & $4.983\pm0.009$ & $0.782\pm0.007$ & $4.8\pm0.2$ & $2.1\pm0.2 $ & $1.5\pm0.2$ & - \tabularnewline
\hline 
\end{tabular}
\par\end{centering}

\protect\caption{{\small{}\label{tab:Details-of-the-FSS}Details of the finite-size
scaling procedure. The left part of the table specifies the data sets
subject to a least squares fit with model \eqref{eq:model} while
the right part gives the fitting results. Numbers with $\pm$ are
error bars (one standard deviation). The range of disorder strength
for all data sets is $K=4.25,\,4.375,\,...\,,\,5.875$. }}
\end{table}

\par\end{center}

\noindent 

\begin{figure}[H]
\begin{centering}
\includegraphics{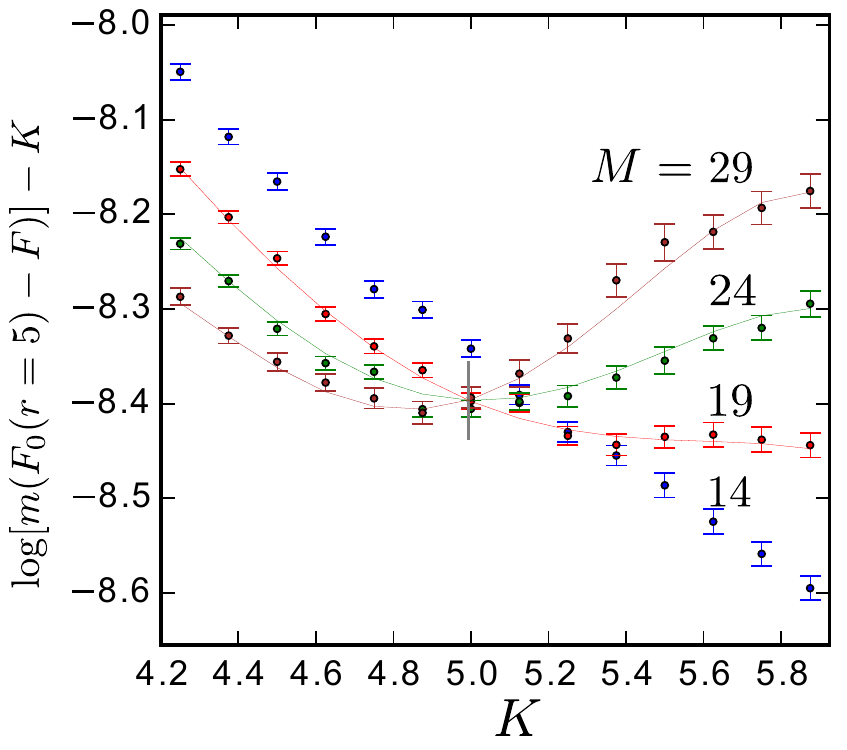}
\par\end{centering}

\protect\caption{{\small{}\label{fig:Fano scaling plot}(color online) 
Scaling
plots for the logarithm of the median $m$ of the distribution of the
difference $\delta F = F_{K=0} - F $ of the two-terminal Fano factor
without and with disorder potential, for a Weyl node with aspect ratio
$r=W/L=5$ and antiperiodic boundarity conditions. For clarity, we plot $log(m)-K$ vs.\ $K$. The
solid curves show the results of a least-squares fit to
a scaling form $m=m_{r}\left(L^{1/\nu}k\right)$ expanded in a fourth order polynomial. The data set for $M=14$ was not included in the analysis. The
estimates for the most important fit parameters are given in 
Table \ref{tab:FanoTable}. The gray vertical
line indicates the position of the estimated critical disorder strength.}}
\end{figure}

\paragraph{Comparison with finite-size scaling for Fano factor.}
As discussed in the main text, besides the conductance, also the Fano factor can be expected to be a suitable observable for a finite-size scaling analysis. A scaling plot is shown in Fig. \ref{fig:Fano scaling plot} and the details of the analysis (done as above for the conductance data) are reported in Table \ref{tab:FanoTable}. Although the number of disorder realizations is comparable to the corresponding conductance data in Fig. \ref{fig:additional scaling plot} (left), for the Fano factor error bars are much larger. Moreover, while for conductance scaling data traces for system sizes $M=14,19,24,29$ all cross in a single point, the Fano factor data for $M=14$ does not cross with the traces of the larger system sizes, indicating that shot noise around criticality is controlled by larger emergent length scales than conductance. For the remaining system sizes, the analysis yields $\nu=1.40\pm0.05$. Given the intrinsic difficulties for the Fano factor data discussed above we consider the error bar overlap with the conductance result $\nu=1.47\pm0.03$ as a confirmation for consistency of the two finite-size scaling methods.  

\begin{center}
\begin{table}[H]
\begin{centering}
\begin{tabular}{|c|c|c|c||c|c|c|c|c|c|c|c|c|}
\hline 
$W/L=r$  & B.C.  & $M$  & $N$  & $\chi^{2}/N$  & $\mathcal{G}$  & $\nu$  & $K_{\rm c}$  & $log(m_\mathrm{c})=a_{0}$  & $a_{1} \cdot 10^1$  & $a_{2} \cdot 10^2$  & $a_{3} \cdot 10^3$ & $a_{4} \cdot 10^3$ \tabularnewline
\hline 
\hline 
5 & APBC   & $19,24,29$  & 42  & 0.72 &  0.69 & $1.40\pm0.05$ & $4.994\pm0.015$ & $-3.4\pm0.01$ & $4.3\pm0.5$ & $5.9\pm1.5$ & $-6\pm2$ & $-5\pm3$    \tabularnewline
\hline 
\end{tabular}
\par\end{centering}
\protect\caption{{\small{}\label{tab:FanoTable}Details of the finite-size
scaling procedure for the Fano factor. The left part of the table specifies the data sets
subject to a least squares fit with model \eqref{eq:model} while
the right part gives the fitting results. Numbers with $\pm$ are
error bars (one standard deviation). The range of disorder strength
 is $K=4.25,\,4.375,\,...\,,\,5.875$. }}
\end{table}

\par\end{center}

\noindent 

\begin{figure}[H]
\begin{centering}
\includegraphics[scale=0.9]{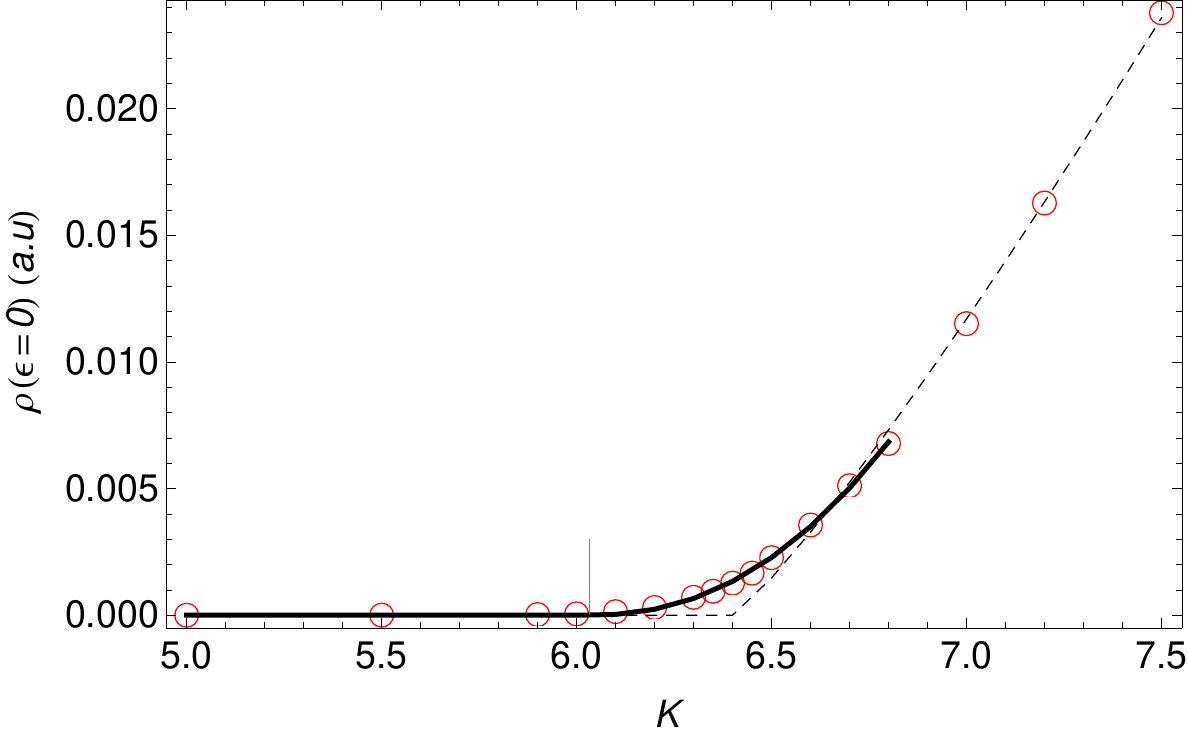}
\par\end{centering}

\protect\caption{{\small{}\label{fig:KobayashiComparison}(color online) 
Bulk density of states of a disordered tight-binding model of a Dirac semimetal at zero energy as a function of disorder strength around criticality. The data (dots) is adapted from Ref. \cite{Kobayashi2014}, Fig. 3a. The solid line is a fit to the scaling form  $\rho(\epsilon=0)\propto (K-K_\mathrm{c})^\beta$ with  $\beta=(3-z) \nu$ fixed to $2.22$ from our finite-size scaling analysis taking into account data points with $K<7$ only. The vertical gray line shows the corresponding estimate  $K_\mathrm{c}=6.03$. The dashed line is a fit of the eight data points with the largest $K$ with setting $K_\mathrm{c}=6.4$ \cite{Kobayashi2014}, the estimate for the exponent is $\beta=1.16$. }}
\end{figure}

\paragraph{Comparison with density-of-states scaling for tight-binding model.}
We revisit the results of a recent density-of-states simulation in a disordered Dirac semimetal from Ref. \cite{Kobayashi2014}. The study is based on a large four-band tight-binding model tuned at the topological phase transition between a strong and weak topological insulator. If inter-node processes can be neglected, around criticality the density of states at zero energy should increase as $\rho(\epsilon=0)\propto (K-K_\mathrm{c})^\beta$ with  $\beta=(3-z) \nu$. Using our estimate $\beta=2.2$, we successfully fit the data from from Ref. \cite{Kobayashi2014},  (cf. Fig. 3a) in Fig. \ref{fig:KobayashiComparison} (solid line), except for the three data points with largest disorder strength. In contrast, the emphasis in the interpretation of Ref. \cite{Kobayashi2014} was laid on data points for larger $K$, excluding the immediate vicinity of the critical point at $K=K_\mathrm{c}$. This leads to a larger estimate for $K_\mathrm{c}$ and a smaller estimate for $\beta$. 


\noindent 

\end{document}